# Monitoring Galvanic Replacement Through Three-Dimensional Morphological and Chemical Mapping

Bart Goris,[1,†] Lakshminarayana Polavarapu,[2,†] Sara Bals,[*,1] Gustaaf Van Tendeloo,[1] Luis M. Liz-Marzán[*,2,3]

**Affiliations:**

[1]EMAT, University of Antwerp, Groenenborgerlaan 171,B-2020 Antwerp, Belgium

[2]Bionanoplasmonics Laboratory, CIC biomaGUNE, Paseo de Miramón 182, 20009 Donostia - San Sebastian, Spain

[3]Ikerbasque, Basque Foundation for Science, 48011 Bilbao, Spain,

[†] These authors contributed equally to this work.

*Correspondence to: sara.bals@ua.ac.be, llizmarzan@cicbiomagunue.es

**Galvanic replacement reactions on metal nanoparticles are often used for the preparation of hollow nanostructures with tunable porosity and chemical composition, leading to tailored optical and catalytic properties. However, the precise interplay between the three-dimensional (3D) morphology and chemical composition of nanostructures during Galvanic replacement is not always well understood as the 3D chemical imaging of nanoscale materials is still challenging. It is especially far from straightforward to obtain detailed information from the inside of hollow nanostructures using electron microscopy techniques such as SEM or TEM. We demonstrate here that a combination of state-of-the-art EDX mapping with electron tomography results in the unambiguous determination of both morphology transformation and elemental composition of nanostructures in 3D, during Galvanic replacement of Ag nanocubes. This work provides direct and unambiguous experimental evidence leading to new insights in the understanding of the galvanic replacement reaction. In addition, the powerful approach presented here can be applied to a wide range of nanoscale transformation processes, which will undoubtedly guide the development of novel nanostructures.**

**One sentence summary:** Three-dimensional morphology and elemental mapping provide completely new insights into the mechanism of Galvanic replacement in silver nanocubes.

Metal nanostructures with well-defined morphologies are of current major relevance because of their multiple potential applications in various fields such as catalysis, plasmonics, photovoltaics, optoelectronics, sensing, drug delivery, bioimaging and therapy[1-6]. Morphology (size and shape) control is crucial for the preparation of nanostructures with specific properties, which in turn determine their ultimate applications.[7-9] A widely employed strategy in the synthesis of nanostructures is based on shape transformations of nanoparticles, either by favoring growth on specific crystalline facets or by etching of the parent nanocrystals.[7,10-13] Among a large number of possibilities, Galvanic replacement reactions and the Kirkendall effect stand out as simple and versatile approaches to transform solid nanoparticles of different shapes into the corresponding porous/hollow nanostructures, where the pore/hole size largely determines their properties.[14,15] Galvanic replacement is a redox process, in which one metal gets oxidized by ions of another metal having a higher reduction potential.[14,16] The oxidized metal thus acts as a sacrificial template for the formation of hollow nanostructures through Galvanic exchange.[16-19] In the case of Kirkendall growth, the difference in diffusion rates of two different metals leads to the formation of voids at the interface and such voids eventually coalesce at the center of the nanoparticles leading to the formation of hollow nanostructures.[15,20] In these processes, the obtained final cage-like nanostructures typically preserve the morphology of the original sacrificial templates. For example, Ag nanocubes are typically converted into either hollow Au-Ag nanocubes or cubic nanoframes.[19,21] However, in rare studies, deviation from the parent template morphology has been observed,[22-24] and such morphology changes may be very promising as they can lead to the development of new complex nanostructures. Both the morphology and elemental composition of nanostructures at different stages of the transformation process have been widely investigated by TEM,[10,15-17,19,23] but a thorough understanding of the different steps in the nanoscale morphological and chemical transformations is currently still lacking.[10,17,25] A major reason is that conventional TEM only yields 2D projection images of these 3D nanostructures, and the outcome of such an analysis may therefore be incomplete or misleading even when the 2D images are acquired with at high resolution. A reliable TEM characterization of nanoparticles should therefore be carried out in 3D. Furthermore, an unambiguous description of the growth process, in particular for hollow nanostructures, ideally requires the simultaneous characterization of both morphology and chemical composition in 3D. Recent developments in the field of electron microscopy have enabled the morphological characterization of nanostructures in 3D by electron tomography.[26-29]. It was shown that by combining a limited number of HAADF-STEM images with a reconstruction algorithm based on compressive sensing, 3D characterization of metal nanoparticles is possible with atomic resolution.[27] A similar approach was furthermore used to investigate the metal-metal interface in an individual Ag@Au nanorod.[30] These results demonstrate the feasibility of chemically sensitive 3D reconstructions with atomic scale resolution. However, such experiments are experimentally and computationally still far from straightforward and very time consuming. Therefore, results can currently only be obtained for model-like systems. It remains especially challenging to resolve chemical composition of more complex nanostructures in 3D. Although the intensity in HAADF-STEM images scales with the atomic number Z of the elements present in the sample, a quantitative interpretation on an absolute scale is still very difficult. In order to investigate the morphology and composition of the hollow particles in this study, 3D HAADF-STEM is insufficient and thus other analytical TEM techniques need to be expanded to 3D. Early attempts to obtain 3D chemical information by energy dispersive X-ray spectroscopy (EDX) were complicated because of the detector inefficiency and the directionality of the sample-detector configuration.[31] Although the Galvanic replacement reaction has been well studied for various plasmonic nanostructures, the current understanding of the reaction mechanism is limited and perhaps speculative, since it is only based on conventional 2D characterization techniques.[7,16,17,23] In this study, we overcome this limitation by applying state of the art (analytical) electron microscopy in 3D to directly monitor the unexpected high yield shape and chemical transformation of Ag nanocubes into single crystalline octahedral Au-Ag nanocages. This process occurs upon a Galvanic replacement reaction between oleylamine-capped Ag nanocubes and $HAuCl_4$ in an organic solvent. Detailed analysis revealed a gradual transformation of Ag nanocubes into Au-Ag octahedral nanocages where all eight corners of the initial nanocubes have transformed into {111} facets of Au-Ag octahedral nanocages, regardless of the degree of truncation of the initial Ag nanocubes. Additionally, 3D-elemental analysis was performed by combining advanced EDX mapping with electron tomography at different stages of Galvanic replacement. These experiments enabled us to directly observe the deposition and diffusion of added Au into the initial Ag nanocubes.

Single crystalline Ag nanocubes were prepared by using oleylamine as both reducing and capping agent, in a Cl-containing organic solvent (details of the synthesis are provided as supplementary materials).[24] The obtained nanocubes were nearly monodisperse with an average edge length of ~24 nm (standard deviation of 6.25%). The transformation of Au nanocubes into octahedral Au-Ag nanocages was performed via titration of the Ag nanocubes colloid with $HAuCl_4$ solution in toluene at 100 ºC (see supplementary materials for details) and initially monitored by UV-visible spectroscopy and TEM characterization (**Figs.S1&S2**). The effects of Galvanic replacement are readily visible through the gradual color changes of the solution, as the amount of added $HAuCl_4$ is increased, which are also reflected in the gradual redshift of the plasmon bands in the UV-visible extinction spectra (**Fig.S1**). During this chemical reaction, the deposition of one gold atom requires the oxidation of 3 silver atoms. TEM suggests that, at the initial stage of the reaction, a single void arises at one of the side facets, and it gradually enlarges when increasing amounts of $HAuCl_4$ are added (**Fig.S2a-d**).This process is thus similar to the well-studied Galvanic reaction on PVP capped Ag nanocubes in aqueous solution, in which hollow cubic nanoboxes are obtained.[17,32] However, when further adding gold salt, hollow particles with apparently different shapes and multiple voids are observed by TEM. Since the reaction is expected to occur similarly on all particles, the different shapes are expected to be related to different orientations of the particles on the TEM grid (**Fig.S2e-f**). Although it is difficult to identify the morphology of the nanoparticles from 2D TEM images, it appears to deviate from the initial cubic structure, which is rather unexpected according to previous reports. At the last reaction stage, the particles on the TEM grid appear to have a 6-fold symmetry with a faceted isotropic shape and an internal void in the middle. The bright field TEM image presented in **Fig. 1** also shows 6 symmetrical features at the edges of each particle. It must be noted that these particles are indeed completely different to the nanocages that were previously obtained from PVP-capped Ag nanocubes, for which cubic nanocages with irregular holes on each facet have been consistently reported.[17] It must be noted that the transformation process was hereby discussed using conventional 2D electron microscopy images. With the aim to characterize the 3D morphology and therefore to understand the growth process of the obtained octahedral nanocages, we performed HAADF-STEM electron tomography. A tilt series of HAADF-STEM projection images was acquired (**Fig.S3**) and used as an input for 3D reconstruction. The result of this process reveals that the obtained hollow nanocages exhibit an almost perfect octahedral morphology with circular holes on the center of every {111} facet, as shown in **Fig.1b** (see **movie S1** for an animated 3D visualization of a selected octahedral nanocage). 3D visualizations of the reconstruction along different orientations are shown in **Fig. 1c-f**. This example clearly demonstrates that 2D TEM images are insufficient to interpret the complex 3D structure of the nanoparticles. In this particular case, the 3D reconstruction enables us to conclude that most particles on the TEM grid (**Fig. 1a**) are oriented along a [111] zone axis and therefore are lying flat on one of their side facets. Even if different particles would have been oriented randomly with respect to the TEM grid, it is still impossible to determine the 3D morphology of a single nanoparticle from such 2D images and small morphological details that could play an important role during the transformation process cannot be detected. **Fig.1g,h** shows two high resolution (HR) HAADF-STEM images of a nanocage (see **Fig. S4** for an image at higher magnification), oriented along two different directions. These images demonstrate that the nanocages retain the monocrystalline character of the original Ag nanocubes. Fourier transform patterns are also shown, which indicate that the major facets correspond to the {111} family from the fcc crystal lattice whereas a small truncation is still observed in the [200] direction **(insets of Fig.1g,h)**. The obtained octahedral nanocages remain stable for extended periods of time (several months at least), but addition of excess gold chloride leads to fragmentation (**Fig.S2h**), which is also reflected in a blue shift of the surface plasmon resonance band (**Fig.S1**).

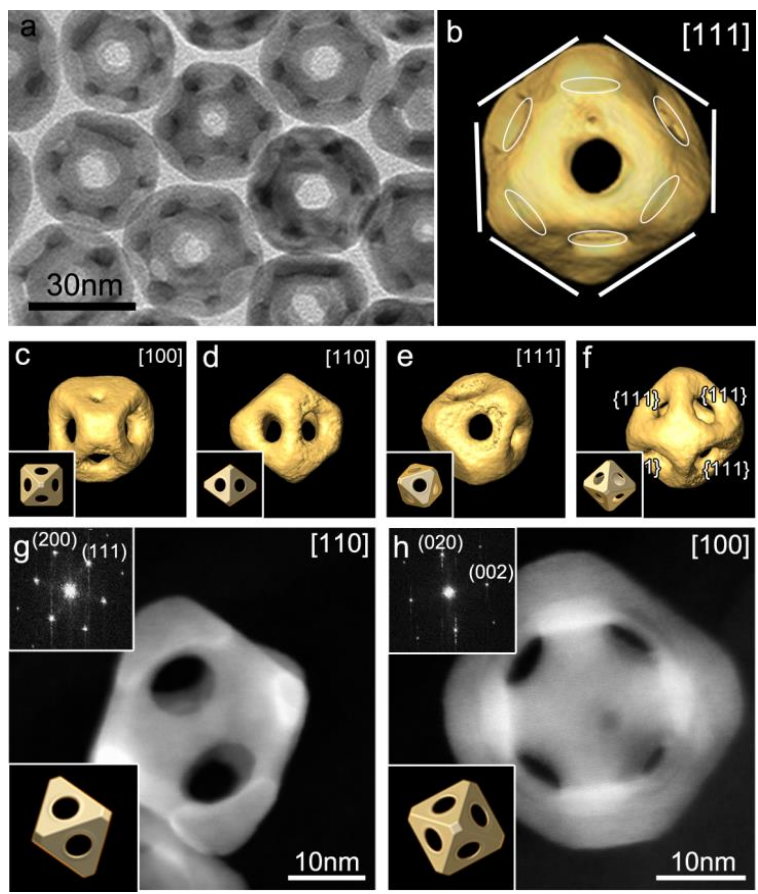

**Fig.1.** (a) BF-TEM image of octahedral nanocages. (b) Electron tomography reconstruction of an octahedral nanocage lying on a {111} facet (see **movie S1** for 3D visualization). (c-f) 3D rendering of octahedral nanocages viewed along different orientations. The insets show 3D models along the corresponding orientation. (g,h) HR-STEM images of an octahedral nanocage along either a [110] (g) or a [100] (h) zone axis orientation (the corresponding models and Fourier transforms are shown as insets).

It has been reported that small changes in the reaction conditions of Galvanic replacement on Ag nanocubes could lead to different types of hollow nanostructures.[19] For instance, the Galvanic replacement reaction on PVP-capped Ag nanocubes leads to cubic nanoboxes and nanocages,[17] but in the presence of CTAB, double walled cubic nanoboxes have been reported as a result of Galvanic replacement and Kirkendall effect at room temperature.[19] For comparison, we also applied electron tomography to nanocages produced from PVP-capped Ag nanocubes, which confirmed that the nanocages preserve the cubic morphology but with some tip truncation forming small {111} faces **(Fig.S5)**. It is interesting to see that the holes are located on these (tiny) {111} faces, just like in the present case.[33] The main difference with the reaction we are reporting thus lies on the presence of oleylamine, which acts as a complexing agent to achieve the dissolution of gold ions in an organic solvent. We therefore postulate that these chemical differences determine the observed morphological changes during the formation of hollow nanoparticles. Such chemical differences very likely imply that the reaction proceeds through a different mechanism, so that the distribution of gold and silver atoms within the particles at various reaction stages may also completely vary. We carefully investigated the transformation of Ag nanocubes into Au-Ag octahedral nanocages using a combination of HAAF-STEM tomography and 3D EDX elemental mapping. This approach results in a powerful 3D characterization of the structure and the precise composition of the NPs at different stages of the reaction.

**Fig.2** presents the morphological evolution from the initial Ag cubes through the final Au/Ag octahedrons by addition of increasing amounts of $HAuCl_4$ solution. For each transformation step, a HAADF-STEM projection is shown that was acquired along a [100] zone axis, together with the corresponding 3D renderings of the reconstructions visualized along different viewing directions and the corresponding models shown as insets, Only from the 3D visualizations we can detect that at the initial stage of the reaction, a small pinhole forms at one of the six side facets of the cube. This hole is indicated by an arrow in **Fig.2a**. This is the first experimental evidence for the existence of the hole on just one of the facets, which had only been postulated on the basis of statistical measurements of TEM/SEM images (*17*). Instead, electron tomography allows us to visualize the formation of the hole together with the porosity of the cube, which is not possible by using either conventional TEM or SEM alone. Such information may allow researchers to accurately investigate the optical and catalytic properties of nanoparticles as a function of their pore size. In addition to the formation of a pinhole, a gradual transition from cubic to octahedral morphology is observed, in such a way that all 8 corners of the initial cube get gradually truncated into 8 octahedron facets. It must be noted that the {100} facets of the starting solid nanocubes have almost completely disappeared in the final structures. Animated movies of each 3D tomographic reconstruction are included as supplementary materials (**movies S2-S4**). Interestingly, 3D investigation of nanocages obtained by Galvanic replacement of a sample containing highly truncated Ag nanocubes shows that the resulting nanocages exhibit the same octahedral morphology (see **Fig. S6**) as those formed from non-truncated cubes. Although it has been reported that the Galvanic reaction on Ag nanocubes initiates at truncated corners,[32] our 3D analysis clearly shows that, under the present conditions, even on highly truncated nanocubes the reaction initiates on a {100} facet, which is similar to the process observed for non-truncated nanocubes (**Fig.S6**).

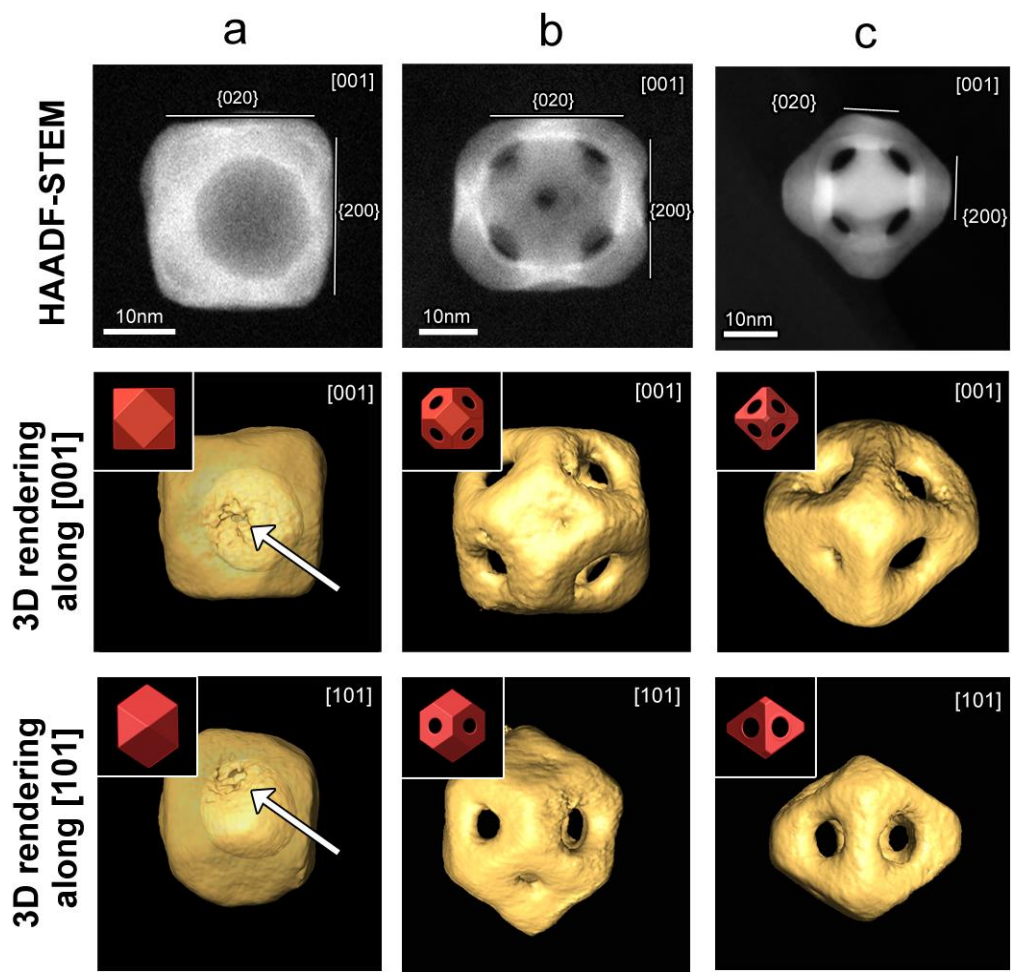

**Fig.2. Transformation of Ag nanocubes into Au-Ag hollow nanocages.** HAADF-STEM images, reconstructed 3D rendering along [001] and [101] directions, and models (insets) showing the formation of hollow nanoparticles after addition of increasing amounts of $HAuCl_4$ from (**a**) through (**c**). An arrow in (**a**) indicates the presence of a pinhole in (only) one of the side facets of the cube during the initial stages of the reaction, from which a void originates.

The intensities in HAADF-STEM images are related to both the thickness and the atomic number Z of the elements present in the sample. Therefore, the 3D reconstructions obtained from HAADF-STEM images not only provide information about the morphology, but also yield qualitative information regarding the distribution of different elements within the nanoparticles. However, precise determination of the elemental distribution at each intermediate step requires recording EDX maps, in particular when alloying is expected. The 2D EDX maps (**Fig.3**) suggest that a thin Au shell is initially deposited at the surface of the Ag nanocubes, which is facilitated by the same crystalline structure and very close lattice constants for both metals. From these 2D EDX maps, it is clear that a hollow structure is present, but no further information can be derived concerning the role of the pinhole on the nanocubes during the chemical transformation. When adding more $HAuCl_4$, it can be seen that the Au and Ag signals overlap in the obtained nanocages (**Fig. 3b**), suggesting that an alloy is formed. For the final stage, the 2D EDX map in **Fig. 3c** shows that an Au-Ag alloy is sandwiched between a thin external Au layer and a thick layer inside the octahedral nanocage. This suggests that gold is deposited inside the hollow nanocubes at the latter stages of the Galvanic replacement reaction. A similar observation has been recently proposed by Puntes and co-workers, regarding the deposition of a Au layer inside hollow cubic double walled nanoboxes by Galvanic replacement. Again, it is important to point out that earlier observations were based on 2D EDX maps only.[19] In addition, the present reaction occurs at 100 ºC, resulting in the faster diffusion of Ag through the Au shells, leading to the formation of the resulting Au/AuAg alloy/Au sandwich structure but with completely different morphology to parent Ag template nanoparticles. Interestingly, the 2D EDX maps of the nanocages obtained from truncated Ag nanocubes (**Fig. S7**) are also similar to the EDX maps of the nanocages obtained from non truncated Ag nanocubes. This indicates that the reaction mechanism seems to be independent of nanocube truncation in the case of oleylamine capped Ag nanocubes in contrast to the well studied Galvanic replacement reaction on PVP capped Ag nanocubes.[17,32]

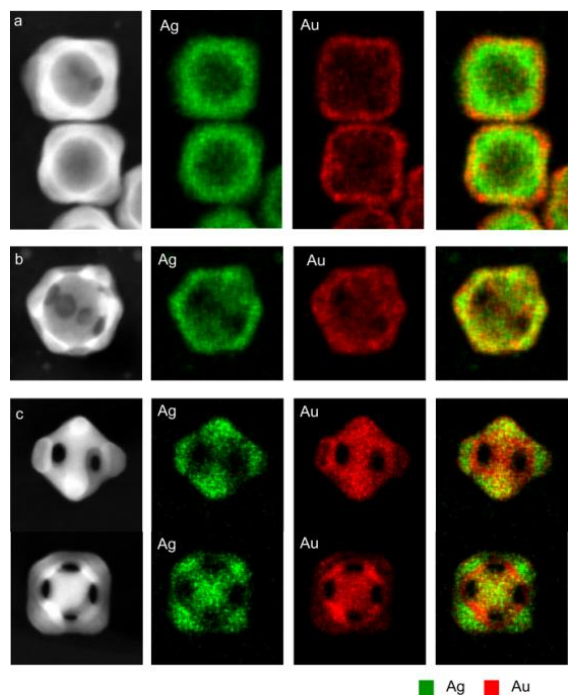

**Fig.3. 2D EDX mapping on nanocages.** For each step in the transformation process with an increasing amount of $HAuCl_4$ (a-c), HAADF-STEM projection images are shown together with the corresponding EDX maps of Au and Ag and the overlaid maps.

Simultaneous investigation of the 3D structural and chemical transformations requires that EDX mapping is combined with electron tomography. In this case, this combined study is expected to confirm the formation of an alloy as well as to elucidate the role of the pinhole. For years, it has been challenging to obtain 3D reconstructions based on EDX projection maps due to the limitations of the specimen-detector configuration.[31] Very recently Arda et al. reported 3D EDX tomography for submicron oxide nanoparticles, but it is far from straightforward to achieve 3D EDX results for smaller nanoparticles.[34] However, recent developments in the design and the high quantum yield of EDX detectors enabled us to obtain 3D EDX mappings of the hollow NPs of such small sizes (~20-30 nm). A tilt series of HAADF-STEM projection images and STEM-EDX maps were acquired simultaneously and used as an input for a 3D reconstruction algorithm. The results are displayed in **Fig. 4** where 3D visualizations along a [100] direction are shown for various representative reaction steps (**Fig. 4a-c**). As discussed above (**Fig. 2**), the evolution of the morphology is especially clear from the HAADF-STEM reconstruction, whereas the corresponding EDX reconstructions yield information on the chemical transition. The 3D EDX results are visualized such that the Au reconstruction is rendered red and Ag corresponds to green (animated views of these reconstructions are available in **supporting movies S5-S7**). To clearly visualize the distribution of the different chemical elements, orthogonal slices through both the HAADF-STEM and the EDX reconstructions are displayed. From these slices, one can indeed clearly and directly observe a thin layer of Au at the beginning of the Galvanic replacement reaction (**Fig. 4a**) which protects the Ag cube from further oxidation. The 3D EDX reconstruction also enables one to investigate the composition at the inside of the particle, which is impossible from 2D EDX maps. It is therefore interesting to note that the 3D EDX reconstruction reveals that a layer of Au is present around the pinhole as shown in **Fig. 5a**, whereas Au is not detected at the inside of the larger void in the cube (see **Fig. 4a**). This direct observation indicates that silver ions diffuse out while the gold atoms are deposited at a lower rate because of the 1:3 stoichiometry as discussed above. Moreover, the gold coating around the pinhole prevents the enlargement of the hole, while allowing silver ions to diffuse out and empty the silver nanocube by etching. When more $HAuCl_4$ was added, the Ag nanocube completely emptied and the gold layer was found to diffuse in and initiate AuAg alloy formation, which causes the reshaping of the hollow NPs, in such a way that lower surface energy {111} facets start to be more pronounced whereas higher surface energy {100} facets decrease in size (**Fig. 2b**). Further addition of $HAuCl_4$ leads to complete reshaping of the nanocages with more pronounced {111} facets and well-defined holes on the center of each facet (**Figs. 2c** and **4c**). The orthoslices through the 3D reconstruction enable a straightforward interpretation of the chemical composition inside the nanostructures and undoubtedly confirm the formation of an alloy at this stage of reaction. In addition, a layer of Au at the inner wall of the void can be observed. Again, it must be noted that such information cannot be obtained in a direct manner from the 3D HAADF-STEM reconstruction and can only be hypothesized in the 2D EDX mapping. It is only by applying EDX in 3D that we can draw reliable and direct conclusions concerning the reactions that take place inside the hollow particles.

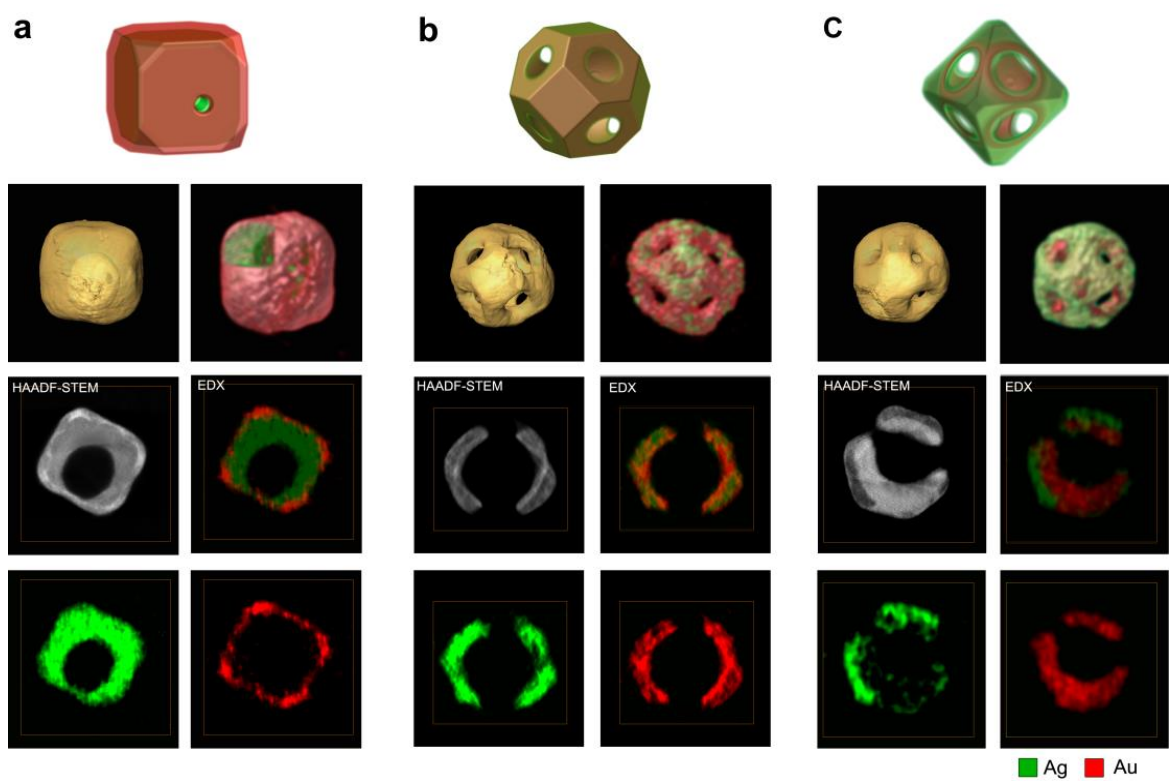

**Fig.4. Visualizations of 3D EDX results.** For each step in the transformation process with increasing amounts of $HAuCl_4$ (a-c), volume rendering viewed along [100] directions are displayed together with 3D rendering of the EDX reconstructions and 3D models (red = Au, green = Ag). Slices through the different reconstructions are also displayed.

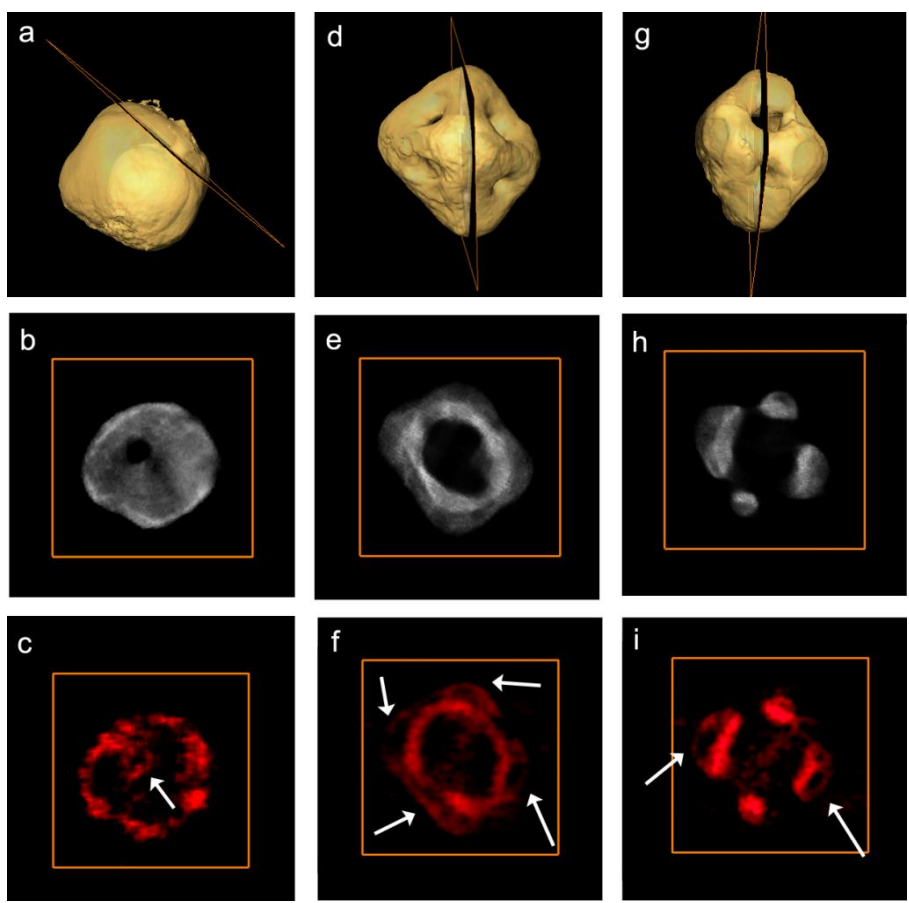

**Fig.5. Au distribution at initial and final stages of Galvanic replacement.** (a,d,g) HAADF-STEM 3D reconstructions. (b-c) Slices *through* the 3D HAADF-STEM reconstruction (b) and 3D EDX reconstruction (c) at the initial stages of the reaction, taken at the position of the pinhole as shown in (a). In (c), a layer of Au is indicated by a white arrow around the pinhole. (d-i) Slices through the 3D HAADF-STEM reconstruction (e,h) and 3D EDX reconstruction (f,i) at the final stage of the reaction taken at different orientations shown in (d) and (g). In (f) and (i), outer layers of Au are indicated by white arrows.

For larger amounts of $HAuCl_4$ solution, {100} facets were found to almost completely disappear as complete transformation of cubes into octahedrons occurred. At this stage, all the initial corners were converted into {111} facets with a central hole (**Fig. 2c**), as clearly depicted in the schematic representation for the gradual shape transformation in **Fig.2**. The results indicate that all the initial {100} facets have been completely etched, thereby reducing the free energy, which in combination with Ostwald ripening and condensation of vacancies into single holes on all {111} facets leads to highly symmetrical octahedral nanocages. The 3D EDX reconstructions indicate that indeed, a thick Au layer is deposited on the inside walls of nanocages after near-complete etching of Ag cube through a pinhole on one of the side facets (**Fig. 4c**). In addition, **Fig. 5d-i** demonstrates that the octahedral nanocages are still covered by a thin outer layer of Au coating. These results indicate that, in addition to the vacancies formed by the Galvanic replacement reaction, Kirkendall growth also contributes to the formation of holes at the latter stages of the reaction. During the Kirkendall growth, as Ag diffuses out faster than the inside diffusion of Au, holes are formed on {111} facets. Interestingly, it has been proposed that this type of Kirkendall growth occurs only at room temperature, leading to double walled nanoboxes, but our results suggest that it can indeed happen at the usual temperature for Galvanic replacement reactions on nanostructures, whereas the final morphology of the parent Ag template is not necessarily retained. Such Au layers inside and outside of the walls protect the nanocages from oxidation and make them unique and stable for months.

In summary, this study shows that combined 3D analysis of morphology and chemical composition is crucial for elucidating complex nanoscale growth processes and provides valuable insights for the better understanding of the Galvanic replacement reaction on nanostructures. Moreover, this method can be applied to understand the exact morphology, growth process and chemical composition of a broad range of nanostructures. We also anticipate that in situ monitoring of the morphological and chemical transformations during Galvanic replacement would lead to even better understanding of the mechanism.[35,36]

**Acknowledgements:**

The authors acknowledge financial support from European Research Council (ERC Advanced Grant # 267867- PLASMAQUO, ERC Advanced Grant # 24691-COUNTATOMS, ERC Starting Grant #335078-COLOURATOMS).The authors also appreciate financial support from the European Union under the Seventh Framework Program (Integrated Infrastructure Initiative N. 262348 European Soft Matter Infrastructure, ESMI).This work was supported by the Flemish Fund for Scientific Research (FWO Vlaanderen) through a PhD research grant to B.G.

**Author contributions:** L.P and L.M.L-M. designed the synthesis experiments and carried out all particles synthesis. B.G. and S.B. performed the 3D tomography, 3D EDX measurements. L.P, B.G, L.M.L-M., S.B. and G.V.T. interpreted the results. All the authors contributed to writing, reading and revising the manuscript.

**Additional information**

Supplementary information is available in the online version of the paper. Reprints and permissions information is available online at www.nature.com/reprints. Correspondence and requests for materials should be addressed to L.M.L-M. or S.B.

**Competing financial interests**

The authors declare no competing financial interests.

# Supplementary Information

# Monitoring Galvanic Replacement through Three-dimensional Morphological and Chemical Mapping

Bart Goris,[1,†] Lakshminarayana Polavarapu,[2,†] Sara Bals,[*,1] Gustaaf Van Tendeloo,[1] Luis M. Liz-Marzán[*,2,3]

[1]EMAT, University of Antwerp, Groenenborgerlaan 171, B-2020 Antwerp, Belgium

[2]Bionanoplasmonics Laboratory, CIC biomaGUNE, Paseo de Miramón 182, 20009 Donostia - San Sebastian, Spain

[3]Ikerbasque, Basque Foundation for Science, 48011 Bilbao, Spain,

[†] These authors contributed equally to this work.

*Correspondence to: sara.bals@ua.ac.be, llizmarzan@cicbiomagunue.es

**This file includes:**

1. Materials and Methods

2. Extinction spectra of nanocubes and nanocages

3. TEM images of nanocubes and nanocages

4. STEM images of octahedral nanocages in different orientations

5. HRTEM image of octahedral nanocage

6. TEM images of PVP capped Ag nanocubes, nanocages and 3D rendering of nanocage.

7. TEM images of oleylamine capped truncated Ag nanocubes, nanocages and their corresponding 3D rendering, EDX images.

Supporting Movie Legends : Movies S1-S7

# Materials and Methods

## Materials: chemicals used for the synthesis

Silver nitrate, silver trifluoroacetate, HAuCl$_4$.3H$_2$O, oleylamine(OA) (technical grade, 70%), 1,2-dichlorobenzene (DCB; 99% extra pure), poly(vinyl pyrrolidone (PVP, MW ≈ 55 000),diethyleneglycol, were purchased from Sigma-Aldrich. All chemicals were used as received.

## Synthesis of silver nanocubes

Silver nanocubes were prepared by previously reported methods with some modifications.[24] In a typical synthesis, 0.39 mmol of AgNO$_3$ and 3 mmol of oleylamine were dissolved in 50 mL (0.44 mol) of DCB and the reaction mixture was sonicated until AgNO$_3$ was completely dissolved. The resulting solution was heated in an oil bath at 165 °C for 12 hours with continuous stirring under atmospheric conditions and then slowly cooled down to room temperature. The resulting solution contained nearly monodisperse silver nanocubes (**Fig.S2a**). Truncated Ag nanocubes were prepared by slightly decreasing the molar ratio of AgNO$_3$ to DCB and prolonging the reaction time. Briefly, 0.3 mmol of AgNO$_3$ and 3 mmol of oleylamine were dissolved in 50 mL (0.44 mol) of DCB by sonication and the resulting solution was heated in an oil bath at 165 °C for 30 hours with continuous stirring under atmospheric conditions to obtain truncated Ag nanocubes. The increased reaction time leads to etching of Ag nanocube edges, as they are more reactive.

## Galvanic replacement reaction

In a typical synthesis, 0.8 mL of as synthesized silver nanocube solution was added to a round-bottom flask containing 20 mL of toluene and the system was heated to 110 °C in an oil bath under magnetic stirring. Different volumes (0.1, 0.15, 0.2, 0.25, 0.3, 0.35, 0.55, 0.75 and 1 mL) of a fresh HAuCl$_4$ stock solution (9 mg of HAuCl$_4$·3H$_2$O in 6 mL of toluene and 0.4 mL of oleylamine) were added to the boiling reaction mixture under magnetic stirring. After each addition the stirring was continued for another 10 min until the UV-Vis spectra did not change further. The obtained nanoparticleswere purified by 1× centrifugation for TEM measurements and at least 3× for tomography and EDX measurements.

**HAADF-STEM imaging**

HAADF-STEM imaging was performed using an aberration corrected cubed FEI Titan operated at 200 kV or 300kV. The probe semiangle of the STEM probe equals 21.4 mrad. During the acquisition of the tomography series, the HAADF detector was positioned at a camera length of 115 mm to avoid coherent imaging.

**EDX mapping**

The EDX maps were acquired with a ChemiSTEM detector. EDX maps were acquired with a screen current of 250 pA and a pixel dwell time of 18 μs. This dwell time results in a frame acquisition time of approximately 2 s after which the drift was corrected using cross correlation. The total acquisition time of each map was 4 minutes. Before using these maps as an input for tomography reconstructions, an averaging filter was used as provided in the Esprit software.

**Three-Dimensional Electron Tomography Reconstructions**

The HAADF-STEM tilt series were acquired with a tilt increment of 2º and a tilt range of approximately ±72º. The EDX maps are recorded with a 5º tilt increment over a range of ±70º. Alignment was performed on the HAADF-STEM tilt series using a cross correlation algorithm in the FEI Inspect3D software and this alignment is then transferred to the EDX tilt series. All reconstructions were performed using the simultaneous iterative reconstruction technique.

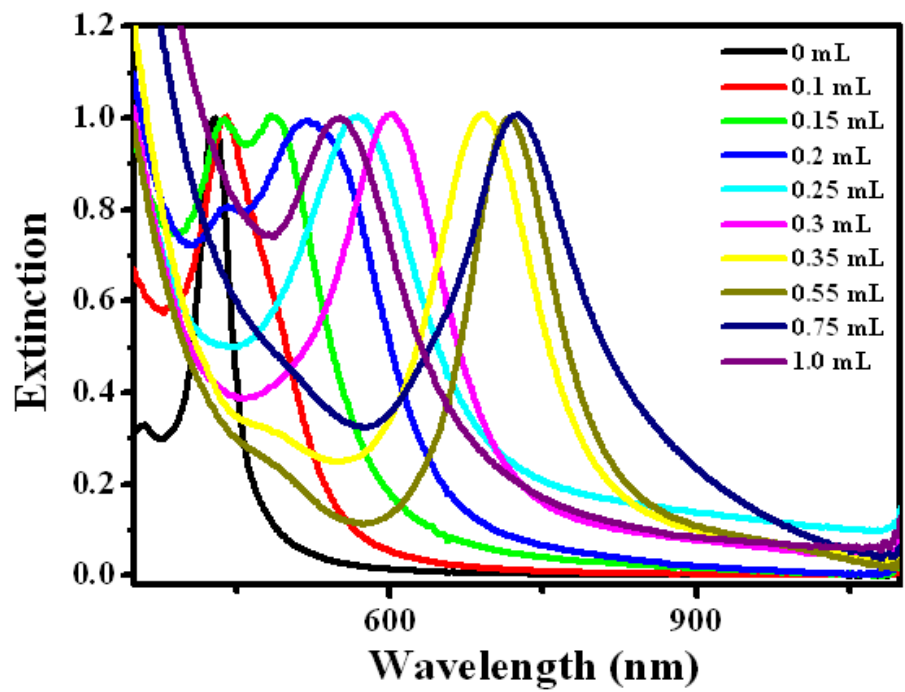

**Fig.S1.** Normalized extinction spectra of nanocube dispersionsbefore (I) and after (II–VIII) Galvanic replacement with different volumes ofHAuCl$_4$: 0.1, 0.15, 0.2, 0.25, 0.3, 0.35, 0.55, 0.75 and 1 mL, respectively.

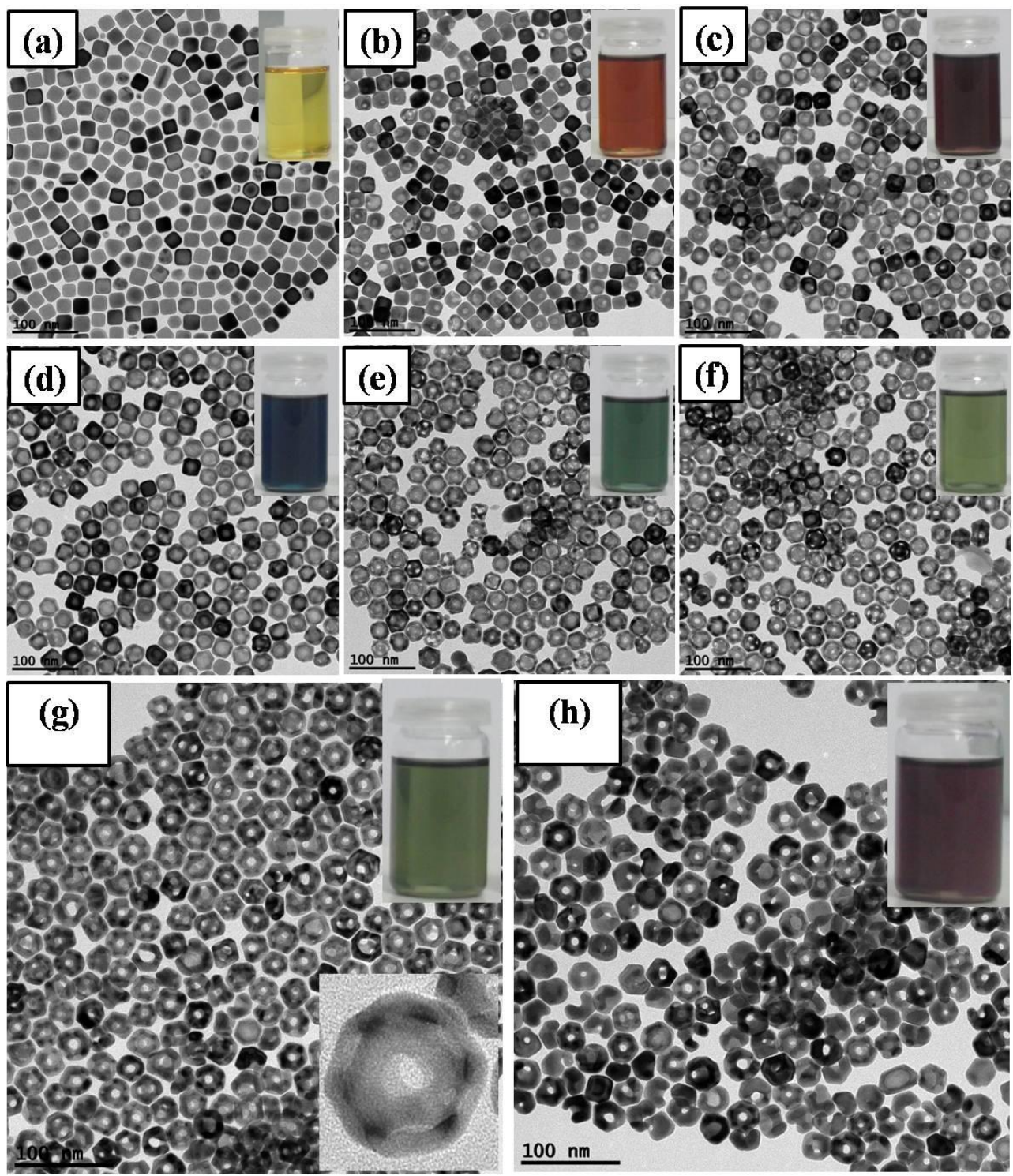

**Fig.S2.** TEM images of theAg nanocubes before (a) and after titration with different volumes of HAuCl₄ solution: (b) 0.1, (c) 0.2, (d) 0.25, (e) 0.35,(f) 0.55, (g) 0.75, (h) 1 mL.

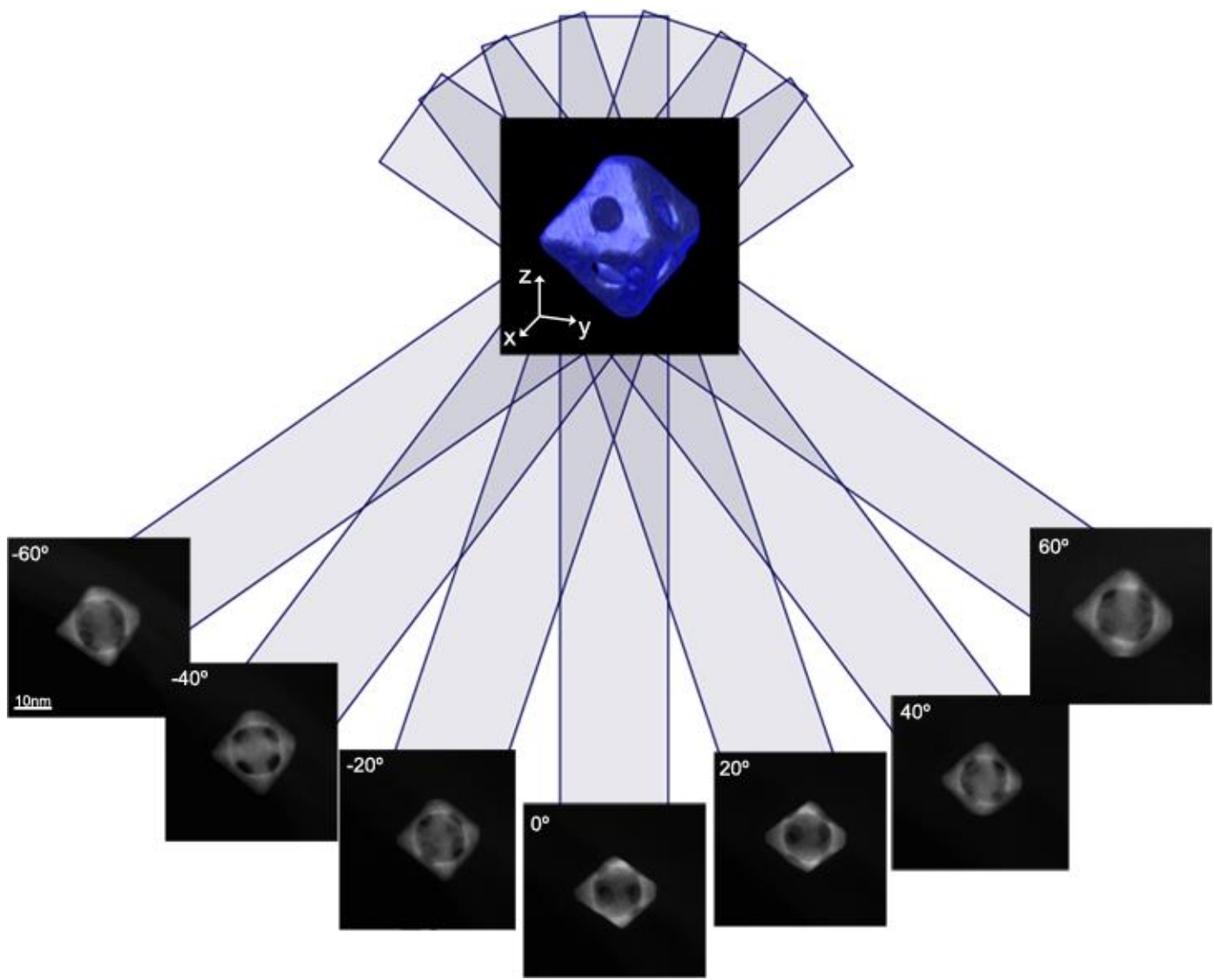

**Fig.S3.** STEM images of an octahedral nanocage acquired at different tilt angles; different morphologies appear for different orientations.

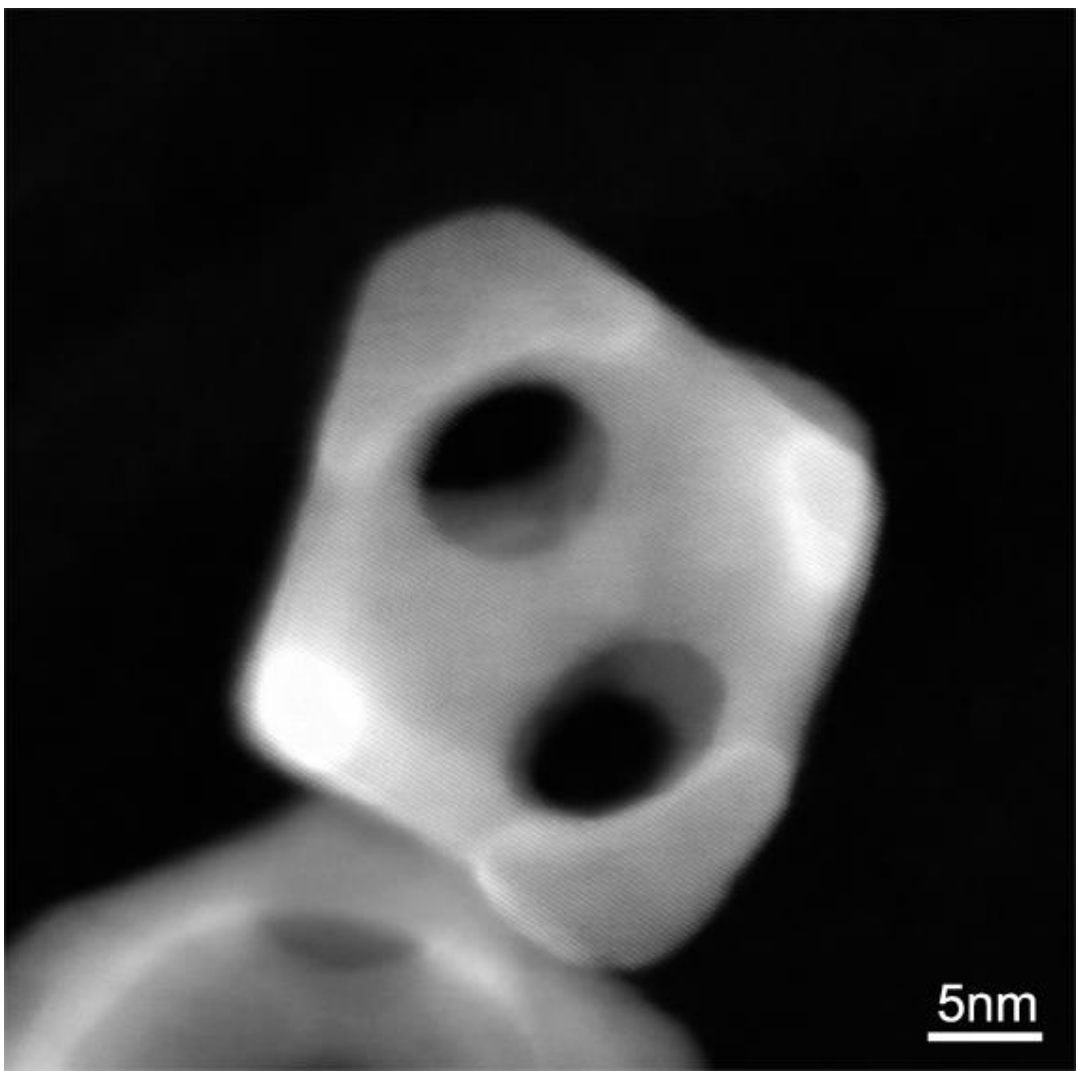

**Fig. S4.** HR-STEM image of anoctahedral nanocage oriented along the [-110] direction. The entire particle appears to be single crystalline without any twinning sites on the crystal.

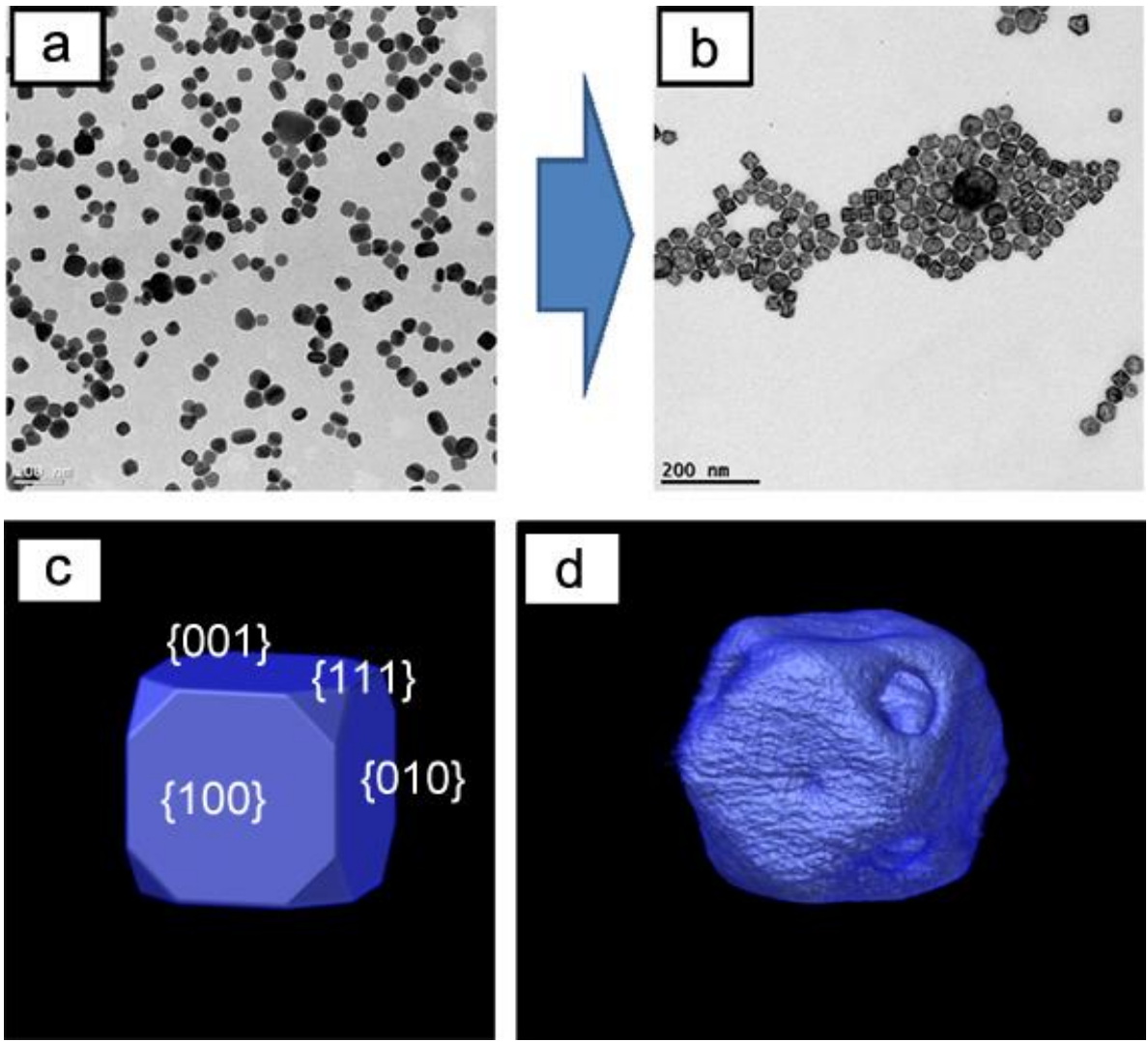

**Fig.S5.** (a) PVP capped Ag nanocubes prepared by the standard polyol method.[37] (b) Au-Ag nanocages obtained after Galvanic replacement. (c-d) Electron tomography reconstruction (d) and model (c) representing the reconstructed imageof thecorrespondingnanocage. The rendering of the 3D reconstruction clearly shows a cubic morphology withslight truncation at the {111} facets. These facets also contain holes.

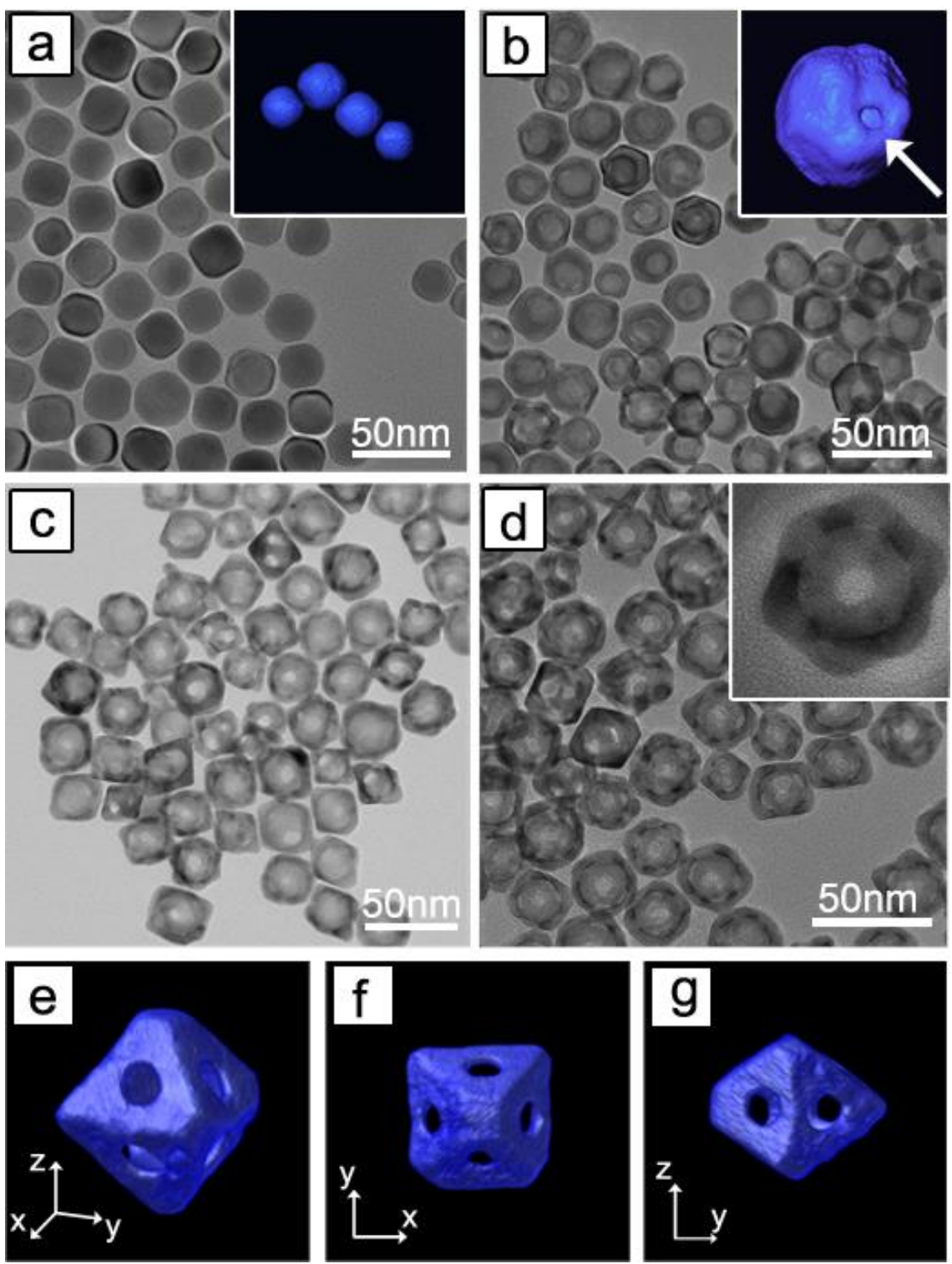

**Fig. S6.** TEM images of truncated Ag nanocubes before (a) and after they reacted with HAuCl$_4$ (b-d). The final particle morphology is similar to that of octahedral particles obtained from Ag nanocubes with sharp edges (e-f). The insets in (a) and (b) show the corresponding 3D reconstructed images, where the cubes clearly appear truncated and the reaction takes place locally on one of the faces; the pinhole is indicated by a white arrow.

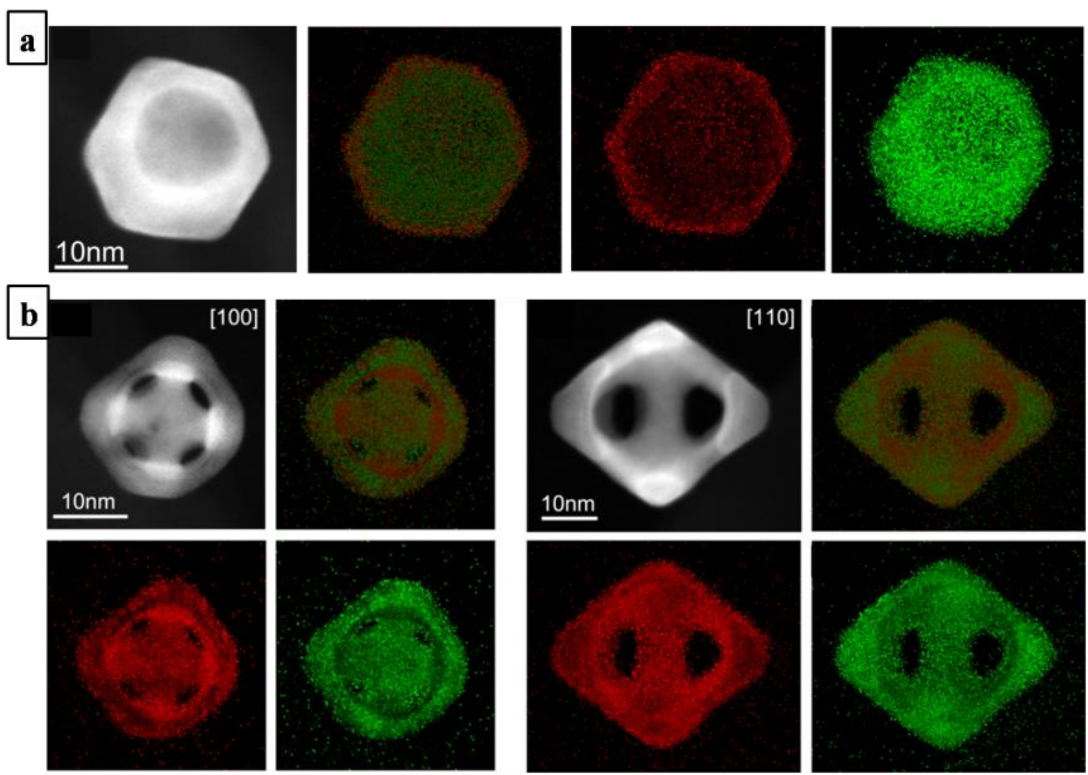

**Fig. S7.** 2D HAADF-STEM projection images and EDX mapping on nanocages. Nanocages obtained at the initial (a) and final (two different orientations) (b) stages of the reaction on truncated Ag nanocubes. These EDX maps are quite similar to the EDX maps of nanocages obtained from non truncated Ag nanocubes, indicating that the reaction mechanism is independent of initial nanocube truncation.

## Supporting Movie Legends

**Movies S1:** Animated view of the tomographic reconstruction of an octahedral nanocage.

**Movies S2-S4:** Animated views of HAADF-STEM 3D reconstructions of nanocages at different stages (**S2:** initial, **S3:** intermediate and **S4**: final) of the galvanic replacement.

**Movies S5-S7:** Animated views of the EDX tomography results at different stages (**S2:** initial, **S3:** intermediate and **S4**: final) of the galvanic replacement.